\documentclass{camera}
\usepackage{graphicx}  

\newcommand{\lesssim}{\lower.5ex\hbox{$\; \buildrel < \over\sim \;$}}
\newcommand{\gtrsim}{\lower.5ex\hbox{$\; \buildrel > \over\sim \;$}}

\begin{document}

%
\title{Highlights from Fermi GRB observations}

%

\author{\vspace{-0.3cm}\\ Jonathan Granot\\ \vspace{0.1cm}
{\rm on behalf of the Fermi LAT and GBM collaborations}}

%
\organization{\vspace{-1.5cm}
Centre for Astrophysics Research, University of Hertfordshire,\\ 
College Lane, Hatfield AL10 9AB, UK}

\maketitle

\vspace{-1.4cm}
\begin{abstract}

The Fermi Gamma-Ray Space Telescope has more than doubled the number
of Gamma-Ray Bursts (GRBs) detected above $100\;$MeV within its first
year of operation. Thanks to the very wide energy range covered by
Fermi's Gamma-ray Burst Monitor (GBM; 8$\;$keV to 40$\;$MeV) and Large
Area Telescope (LAT; 25$\;$MeV to $>\,$300$\;$GeV) it has measured the
prompt GRB emission spectrum over an unprecedentedly large energy
range (from $\sim 8\;$keV to $\sim 30\;$GeV). Here I briefly outline
some highlights from Fermi GRB observations during its first $\sim
1.5\;$yr of operation, focusing on the prompt emission
phase. Interesting new observations are discussed along with some of
their possible implications, including: (i) What can we learn from the
Fermi-LAT GRB detection rate, (ii) A limit on the variation of the
speed of light with photon energy (for the first time beyond the
Planck scale for a linear energy dependence from direct time of
arrival measurements), (iii) Lower-limits on the bulk Lorentz factor
of the GRB outflow (of $\sim 1000$ for the brightest Fermi LAT GRBs),
(iv) The detection (or in other cases, lack thereof) of a distinct
spectral component at high (and sometimes also at low) energies, and
possible implications for the prompt GRB emission mechanism, (v) The
later onset (and longer duration) of the high-energy emission
($>\,$100$\;$MeV), compared to the low-energy ($\lesssim\,$1$\;$MeV)
emission, that is seen in most Fermi-LAT GRBs.

\end{abstract}

%

\section{Pre-Fermi high-energy GRB observations}

The Energetic Gamma-Ray Experiment Telescope (EGRET) on-board the
Compton Gamma-Ray Observatory (CGRO; 1991$-$2000) was the first to
detect high-energy emission from GRBs. EGRET detected only five GRBs
with its Spark Chambers (20$\,$MeV to 30$\,$GeV) and a few GRBs with
its Total Absorption Shower Counter (TASC; $1-200\;$MeV).
Nevertheless, these events already showed diversity. The most
prominent examples are GRB~940217, with high-energy emission lasting
up to $\sim 1.5\;$hr after the GRB including an $18\;$GeV photon after
$\sim 1.3\;$hr,~\cite{Hurley93} and GRB~941017 which had a distinct
high-energy spectral component~\cite{Gonzalez03} detected up to $\sim
200\;$MeV with $\nu F_\nu \propto \nu$. This high-energy spectral
component had $\sim 3$ times more energy and lasted longer ($\sim
200\;$s) than the low-energy (hard X-ray to soft gamma-ray) spectral
component (which lasted several tens of seconds), and may be naturally
explained as inverse-Compton emission from the forward-reverse shock
system that is formed as the ultra-relativistic GRB outflow is
decelerated by the external medium~\cite{GrGu03,PW04}. Nevertheless,
better data are needed in order to determine the origin of such
high-energy spectral components more conclusively. The Italian
experiment Astro-rivelatore Gamma a Immagini LEggero (AGILE; launched
in 2007) has detected GRB~080514B at energies up to $\sim 300\;$MeV,
and the high-energy emission lasted longer ($>\,$13$\;$s) than the
low-energy emission ($\sim 7\;$s)~\cite{Giuliani08}.
Below are some highlights of Fermi GRB observations so far and what
they have taught us.

\section{LAT GRB detection rate: what can it teach us?}

\begin{figure}
\centerline{\includegraphics[width=0.92\textwidth]{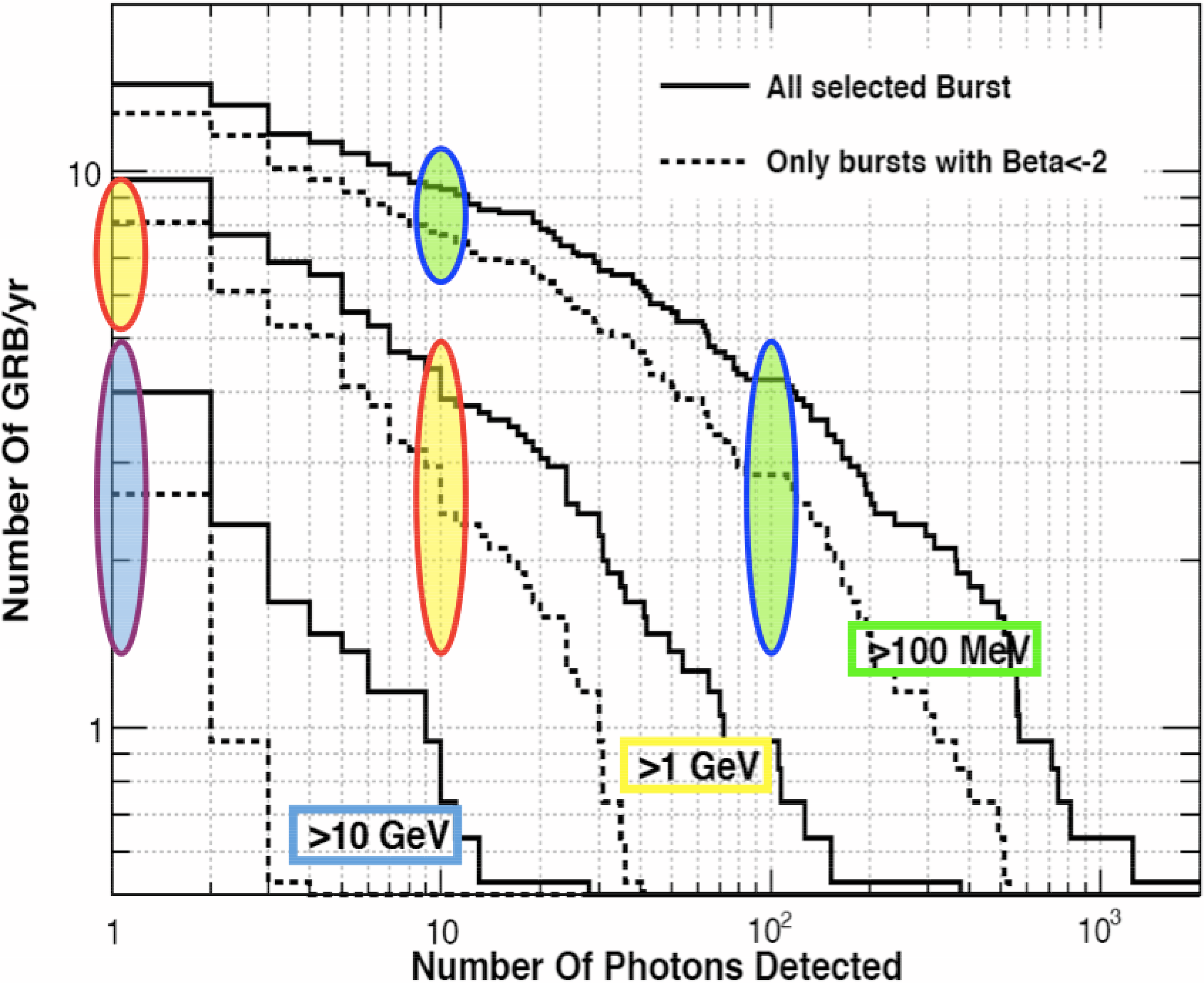}}
\caption{
LAT GRB detection rates (color ellipses) superposed on top of
pre-launch expected rates based on the extrapolation of a Band
spectrum fit from the BATSE energy range~\cite{pre-launch}. The
ellipses' inner color indicates the minimal photon energy (green,
yellow and cyan correspond to 0.1, 1 and 10$\;$GeV, respectively),
while their hight indicates the uncertainty ($\pm N^{1/2}/1.5\,$yr) on
the corresponding LAT detection rate ($N/1.5\,$yr) due to the small
number ($N$) of detected GRBs.}
\label{fig:pre-launch}
\end{figure}

During its first $1.5\;$yr of routine operation, from Aug.~2008 to
Jan.~2010, the LAT has detected 14 GRBs, corresponding to a detection
rate of $\sim 9.3\;{\rm yr^{-1}}$. Table~\ref{tab:LAT14GRBs}
summarizes their main properties. While at least 13 of the 14 LAT GRBs
had $\geq 10$ photons above $100\;$MeV, 4 were particularly bright in
the LAT with $\geq 1$ photon above $10\;$GeV, $\geq 10$ photons above
$1\;$GeV, and $\geq 100$ photons above $100\;$MeV. This corresponds to
a bright LAT GRB (as defined above) detection rate of $\sim
2.7\;$GRB/yr (with a rather large uncertainty due to the small number
statistics). There were also 11 GRBs with $\geq 1$ photon above
$1\;$GeV, corresponding to $\sim 7.3\;$GRB/yr. These detection rates
are compatible with pre-launch expectations~\cite{pre-launch} based on
a sample of bright BATSE GRBs for which the fit to a Band spectrum
over the BATSE energy range ($20\;$keV to $2\;$MeV) was extrapolated
into the LAT energy range (see Fig.~\ref{fig:pre-launch}). The
agreement is slightly better when excluding cases with a rising $\nu
F_\nu$ spectrum at high energies (i.e. a high-energy photon index
$\beta > -2$).~\footnote{Such a hard high-energy photon index may be
an artifact of the limited energy range of the fit to BATSE data, or
may be affected by poor photon statistics at $\gtrsim 1\;$MeV.} This
suggests that, on average, there is no significant excess or deficit
of high-energy emission in the LAT energy range relative to such an
extrapolation from lower energies.  As described in
\S~\ref{sec:delay+HEcomp}, however, in individual cases we do have evidence for
such an excess.  The observed LAT GRB detection rate implies that, on
average, only about $\sim 10-20\%$ of the energy that is radiated
during the prompt GRB emission phase is channeled into the LAT energy
range, suggesting that in most GRBs the high-energy radiative output
does not significantly affect the total energy budget. Short GRBs,
however, appear to be different in this respect (see
\S~\ref{sec:long-vs-short} and Fig.~\ref{fig:long-vs-short}).

\begin{table}
\begin{center}
{\scriptsize
\begin{tabular}{|c|c|c|c|c|c|c|c|c|c|}
\hline
    &                    & long  & \multicolumn{2}{|c|}{number of} &
\multicolumn{2}{|c|}{HE emission} & extra    & highest &  \cr\cline{6-7}

GRB & $\theta_{\rm LAT}$ & or    & \multicolumn{2}{|c|}{events above} &
starts         & lasts            & spec. & energy  & $z$ \cr\cline{4-5}
    &                    & short & 0.1$\,$GeV & 1$\,$GeV &
later          & longer           & comp.    & (GeV)   &
\\
\hline
\hline
080825C & $\sim\,$60$^\circ$ & long  & $\sim\,$10 & 0       & ?   & yes & no  & $\sim\,$0.6 & ----- \\
080916C & $49^\circ$         & long  & 145        & 14      & yes & yes & ?   & $\sim\,$13  & $\sim\,$4.35 \\
081024B & $21^\circ$         & short & $\sim\,$10 & 2       & yes & yes & ?   & $\sim\,$3   & ----- \\
081215A & $\sim\,$86$^\circ$ & long  & -----      & ---     & ?   & ?   & --- & -----       & ----- \\
090217  & $\sim\,$34$^\circ$ & long  & $\sim\,$10 & 0       & no  & no  & no  & $\sim\,$1   & ----- \\
090323  & $\sim\,$55$^\circ$ & long  & $\sim\,$20 & $>\,$0  & ?   & yes & ?   & ?           & 3.57 \\
090328  & $\sim\,$64$^\circ$ & long  & $\sim\,$20 & $>\,$0  & ?   & yes & ?   & ?           & 0.736 \\
090510  & $\sim\,$14$^\circ$ & short & $>\,$150   & $>\,$20 & yes & yes & yes & $\sim\,$31  & 0.903 \\
090626  & $\sim\,$15$^\circ$ & long  & $\sim\,$20 & $>\,$0  & ?   & yes & ?   & ?                      & ----- \\
090902B & $51^\circ$         & long  & $>\,$200   & $>\,$30 & yes & yes & yes & $\sim\,$33  & 1.822 \\
090926  & $\sim\,$52$^\circ$ & long  & $>\,$150   & $>\,$50 & yes & yes & yes & $\sim\,$20  & 2.1062 \\
091003A & $\sim\,$13$^\circ$ & long  & $\sim\,$20 & $>\,$0  & ?   & ?   & ?   & ?                      & 0.8969 \\
091031  & $\sim\,$22$^\circ$ & long  & $\sim\,$20 & $>\,$0  & ?   & ?   & ?   & $\sim\,$1.2 & ----- \\
100116A & $\sim\,$29$^\circ$ & long  & $\sim\,$10 & 3       & ?   & ?   & ?   & $\sim\,$2.2 & ----- \\
\hline 
\end{tabular}}
\end{center}
\caption{\footnotesize
Summary of the 14 GRBs detected by the LAT between August 2008 and
January 2010 -- its first 1.5 years of routine operation following
Fermi's launch on 11 June 2008; $\theta_{\rm LAT}$ is the angle from
the LAT boresight at the time of the GBM GRB trigger.
\label{tab:LAT14GRBs}}
\end{table}

\section{Limits on Lorentz Invariance Violation}
\label{sec:LIV}

Some quantum gravity models allow violation of Lorentz invariance, and
in particular allow the photon propagation speed, $v_{\rm ph}$, to
depend on its energy, $E_{\rm ph}$: $v_{\rm ph}(E_{\rm ph}) \neq c$,
where $c\equiv\displaystyle\lim_{E_{\rm ph}\to 0}v_{\rm ph}(E_{\rm
ph})$. The Lorentz invariance violating (LIV) part in the dependence
of the photon momentum, $p_{\rm ph}$, on its energy, $E_{\rm ph}$, can be
expressed as a power series,
\begin{equation}\label{eq:QG1}
\frac{p_{\rm ph}^2c^2}{E_{\rm ph}^2}-1 
= \sum_{k = 1}^\infty s_k\left(\frac{E_{\rm ph}}{M_{{\rm QG},k}c^2}\right)^k\ ,
\end{equation}
in the ratio of $E_{\rm ph}$ and a typical energy scale $M_{{\rm
QG},k}c^2$ for the $k^{\rm th}$ order, which is expected to be up to
the order of the Planck scale, $M_{\rm Planck} = (\hbar c/G)^{1/2}
\approx 1.22\times 10^{19}\;{\rm GeV/c^2}$, where $s_k \in
\{-1,\,0,\,1\}$. Since we observe photons of energy well below the
Planck scale, the dominant LIV term is associated with the lowest
order non-zero term in the sum, of order $n = \min\{k|s_k\neq 0\}$,
which is usually assumed to be either linear ($n = 1$) or quadratic
($n = 2$). The photon propagation speed is given by the corresponding
group velocity,
\begin{equation}
v_{\rm ph} = \frac{\partial E_{\rm ph}}{\partial p_{\rm ph}} \approx
c\left[1-s_n\,\frac{n+1}{2}\left(\frac{E_{\rm ph}}{M_{{\rm QG},n}c^2}\right)^n\,\right]\ .
\end{equation}
Note that $s_n = 1$ corresponds to the sub-luminal case ($v_{\rm ph} <
c$ and a positive time delay), while $s_n = -1$ corresponds to the
super-luminal case ($v_{\rm ph} > c$ and a negative time delay).
Taking into account cosmological effects~\cite{JP08}, this induces a
time delay (or lag) in the arrival of a high-energy photon of energy
$E_{\rm h}$, compared to a low-energy photon of energy $E_{\rm l}$
(emitted simultaneously at the same location), of
\begin{equation}\label{eq:Dt}
\Delta t = s_n\,\frac{(1+n)}{2H_0}\frac{\left(E_h^n-E_l^n\right)}{(M_{{\rm
QG},n}c^2)^n} \int_{0}^{z}\frac{(1+z^{\prime})^n}
{\sqrt{\Omega_m(1+z^{\prime})^3+\Omega_{\Lambda}}}\,dz^{\prime}\ .
\end{equation}
Here we concentrate on our results for a linear energy dependence ($n
= 1$).  

We have applied this formula to the highest energy photon detected in
GRB~080916C, with an energy of $E_{\rm h} =
13.22^{+0.70}_{-1.54}\;$GeV, which arrived at $t = 16.54\;$s after the
GRB trigger (i.e. the onset of the $E_l \sim 0.1\;$MeV emission).
Since it is hard to associate the highest energy photon with a
particular spike in the low-energy lightcurve, we have made the
conservative assumption that it was emitted anytime after the GRB
trigger, i.e. $\Delta t \leq t$, in order to obtain a limit for the
sub-luminal case ($s_n = 1$): $M_{{\rm QG},1} > 0.1M_{\rm Planck}$.
This was the strictest limit of its kind~\cite{080916C}, at that time.

\begin{figure}
\centerline{\hspace{0.88cm}\includegraphics[width=12.6cm]{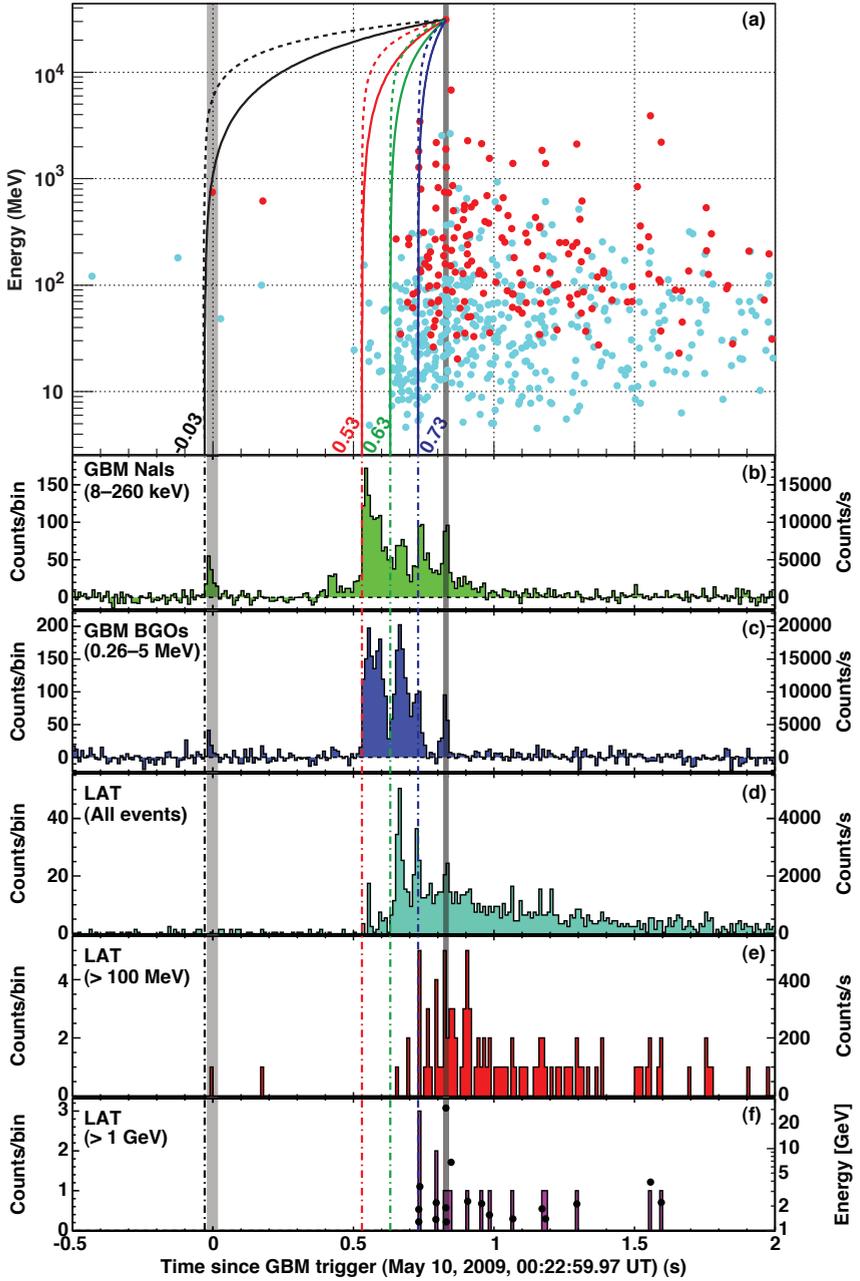}}
\caption{
Light curves of GRB~090510 at different energies 
(for details see~\cite{090510-LIV}).}
\label{fig:090510-LIV}
\end{figure}

However, the next very bright LAT GRB, 090510, was short and had very
narrow sharp spikes in its light curve (see Fig~\ref{fig:090510-LIV}),
thus enabling us to do even better~\cite{090510-LIV}. Our main results
for GRB~090510 are summarized in Table~\ref{tab:LIV}. The first 4
limits are based on a similar method as described above for
GRB~080916C, using the highest energy photon, $E_h =
30.53_{-2.56}^{+5.79}\;$GeV, and assuming that its emission time $t_h$
was after the start of a relevant lower energy emission episode: $t_h
> t_{\rm start}$. These 4 limits correspond to different choices of
$t_{\rm start}$, which are shown by the vertical lines in
Fig.~\ref{fig:090510-LIV}. We conservatively used the low end of the
$1\;\sigma$ confidence interval for the highest energy photon ($E_h =
28\;$GeV) and for the redshift ($z = 0.900$). The most conservative
assumption of this type is associating $t_{\rm start}$ with the onset
of any detectable emission from GRB~090510, namely the start of the
small precursor that GBM triggered on, leading to $\xi_1 = M_{{\rm
QG},1}/M_{\rm Planck} > 1.19$. However, it is highly unlikely that the
$31\;$GeV photon is indeed associated with the small precursor. It is
much more likely associated with the main soft gamma-ray emission,
leading to $\xi_1 > 3.42$. Moreover, for any reasonable emission
spectrum, the emission of the $31\;$GeV photon would be accompanied by
the emission of a large number of lower energy photons, which would
suffer a much smaller time delay due to LIV effects, and would
therefore mark its emission time. We could easily detect such photons
in energies above $100\;$MeV, and therefore the fact that significant
high-energy emission is observed only at later times (see
Fig.~\ref{fig:090510-LIV}) strongly argues that the $31\;$GeV photon
was not emitted before the onset of the observed high-energy
emission. One could choose either the onset time of the emission above
$100\;$MeV or above $1\;$GeV, which correspond to $\xi_1 > 5.12$, and
$\xi_1 > 10.0$, respectively.~\footnote{We note that there is no
evidence for LIV induced energy dispersion that might be expected if
indeed the $31\;$GeV photon was emitted near our choices for $t_{\rm
start}$, together with lower energy photons, as can be expected for
any reasonable emission spectrum. This is evident from the lack of
accumulation of photons along the {\it solid} curves in panel (a) of
Fig.~\ref{fig:090510-LIV}, at least for the first 3 $t_{\rm start}$
values, and provides support for these choices of $t_{\rm start}$
(i.e. that they can indeed serve as upper limits on a LIV induced
energy dispersion).}

The 5$^{\rm th}$ and 6$^{\rm th}$ limits in Table~\ref{tab:LIV} are
more speculative, as they rely on the association of an individual
high-energy photon with a particular spike in the low-energy light
curve, on top of which it arrives. While these associations are not
very secure (the chance probability is roughly $\sim 5-10\%$), they
are still most likely, making the corresponding limits interesting,
while keeping this big caveat in mind. The allowed emission time of
these two high-energy photons, if these associations are real, is
shown by the two thin vertical shaded regions in
Fig.~\ref{fig:090510-LIV}. For the $31\;$GeV photon this gives a limit
of $\xi_1 > 102$ for either sign of $s_n$.

The last limit in Table~\ref{tab:LIV} is based on a different method,
which is complementary and constrains both signs of $s_n$. It relies
on the highly variable high-energy light curve, with sharp narrow
spikes, which would be smeared out if there was too much energy
dispersion (of either sign). We have used the DisCan
method~\cite{DisCan} to search for linear energy dispersion within the
LAT data (the actual energy range of the photons used was $35\;$MeV to
$31\;$GeV)~\footnote{We obtain similar results even if we use only
photons below $3\;$GeV or $1\;$GeV.} during the most intense emission
interval (0.5$\,$--$\,$1.45$\,$s). This approach extracts dispersion
information from all detected LAT photons and does not involve binning
in time or energy. Using this method we obtained a robust lower limit
of $\xi_1 > 1.22$ (at the 99\% confidence level). 

Our most conservative limits (the first and last limits in
Table~~\ref{tab:LIV}) rely on very different and largely independent
analysis, yet still give a very similar limit: $M_{{\rm QG},1} >
1.2M_{\rm Planck}$. This lends considerable support to this result,
and makes it more robust and secure than for each of the methods
separately.

\newcommand{\rb}[1]{\raisebox{1.3ex}[0pt]{#1}}
\newcommand{\lb}[1]{\raisebox{-1.3ex}[0pt]{#1}}
\begin{table}
\begin{center}
{\footnotesize
\begin{tabular}{|c|c|c|c|c|c|}
\hline
$t_{\rm start}$ & limit on          & Reason for choice of                   & 
$E_l$ & valid     & lower limit on  \\ 
(ms)            & $|\Delta t|$ (ms) & $t_{\rm start}$ or limit on $\Delta t$ & 
(MeV) & for $s_n$ & $M_{{\rm QG},1}/M_{\rm Planck}$
\\
\hline
\hline
$-30$ & $<859$ & {\scriptsize start of any observed emission}  & 
0.1  & 1 & ${\bf >1.19}$ \\
$530$ & $<299$ & {\scriptsize start of main $<\,$1$\,$MeV emission} & 
0.1  & 1 & ${\bf >3.42}$ \\
$630$ & $<199$ & {\scriptsize start of $>100\;$MeV emission}    & 
100  & 1 & ${\bf >5.12}$ \\
$730$ &  $<99$ & {\scriptsize start of $>1\;$GeV emission}      & 
1000 & 1 & ${\bf >10.0}$ \\
\hline
 ---  &  $<10$ & {\scriptsize association with $<\,$1$\,$MeV spike} & 
0.1  & $\pm\,$1 & ${\bf >102}$  \\
 ---  &  $<19$ & {\scriptsize if $0.75\,$GeV $\gamma$ is from $1^{\rm st}$ spike} & 
0.1  & $\pm\,$1 & ${\bf >1.33}$  \\ \hline
\multicolumn{2}{|c|}{$|\frac{\Delta t^{^{}}}{\Delta E_{}}| < 30\;\frac{\rm ms}{\rm GeV}$} & 
lag analysis of all LAT events &
--- & $\pm\,$1 & ${\bf >1.22}$
\\
\hline 
\end{tabular}}
\end{center}
\caption{\footnotesize
Lower-limits on the Quantum Gravity (QG) mass scale associated with a
possible linear ($n=1$) variation of the speed of light with photon
energy, that we can place from the lack of time delay (of sign $s_n$)
in the arrival of high-energy photons relative to low-energy photons,
from our observations of GRB~090510 (from~\cite{090510-LIV}).
\label{tab:LIV}}
\end{table}

\section{Lower limits on the bulk Lorentz factor}
\label{sec:Gamma_min}

The GRB prompt emission typically has very large isotropic equivalent
luminosities ($L \sim 10^{50}-10^{53}\;{\rm erg\;s^{-1}}$),
significant short time scale variability, and typical photon energies
$\gtrsim m_ec^2$ (in the source cosmological frame). This would result
in a huge optical depth to pair production ($\gamma\gamma \to e^+e^-$)
at the source, which would thermalize the spectrum and thus be at odds
with the observed non-thermal spectrum, unless the emitting material
was moving toward us relativistically, with a bulk Lorentz factor
$\Gamma\gg 1$. This ``compactness'' argument has been used to derive
lower-limits, $\Gamma_{\rm min}$, on the value of $\Gamma$, which were
typically $\sim 10^2$ and in some cases as high as a few hundred
(see~\cite{LS01} and references therein). However, the photons that
provided the opacity for these limits were well above the observed
energy range, so there was no direct evidence that they actually
existed in the first place.

With Fermi, however, we adopt a more conservative approach of relying
only on photons within the observed energy range. Under this approach,
\begin{equation}
\Gamma_{\rm min} \lesssim (1+z)\frac{E_{\rm ph,max}}{m_ec^2} 
\approx 200(1+z)\left(\frac{E_{\rm ph,max}}{100\,{\rm MeV}}\right)\ , 
\end{equation}
where $E_{\rm ph,max}$ is the highest observed photon energy, so that
setting a large $\Gamma_{\rm min}$ requires observing sufficiently
high-energy photons.

The main uncertainty in deriving $\Gamma_{\rm min}$ is usually the
exact choice for the variability time, $t_v$. Other uncertainties
arise from those on the spectral fit parameters, or on the degree of
space-time overlap between the high-energy photon and lower energy
target photons, in cases where there is more than one spectral
component without conclusive temporal correlation between their
respective light curves. Finally, the fact that our limits rely on a
single high-energy photon also induces an uncertainty, as it might
still escape from an optical depth of up to a few. However, in most
cases the second or third highest-energy photons help to relax the
affect this has on $\Gamma_{\rm min}$ (as the probability that
multiple photons escape from $\tau_{\gamma\gamma} > 1$ rapidly
decreases with the number of photons). Thus, we have derived
reasonably conservative $\Gamma_{\rm min}$ values for 3 of the
brightest LAT GRBs: $\Gamma_{\rm min}
\approx 900$ for GRB~080916C~\cite{080916C}, $\Gamma_{\rm min} \approx
1200$ for GRB~090510~\cite{090510-phys}, and $\Gamma_{\rm min} \approx
1000$ for GRB~090902B~\cite{090902B}.  This shows that short GRBs
(such as 090510) are as highly relativistic as long GRBs (such as
080916C or 090902B), which was questioned before the launch of
Fermi~\cite{Nakar07}. Since our highest values of $\Gamma_{\rm min}$
are derived for the brightest LAT GRBs, they are susceptible to strong
selection effects. It might be that GRBs with higher $\Gamma$ tend to
be brighter in the LAT energy range (e.g. by avoiding intrinsic pair
production~\cite{Ghisellini10}).

\section{Delayed onset and a distinct high-energy spectral component}
\label{sec:delay+HEcomp}

A delayed onset of the high-energy emission ($>\,$100$\;$MeV) relative
to the low-energy emission ($\lesssim 1\;$MeV) appears to be a very
common feature in LAT GRBs. It clearly appears in all 4 of the
particularly bright LAT GRBs, while in dimmer LAT GRBs it is often
inconclusive due to poor photon statistics near the onset time.  The
time delay, $t_{\rm delay}$, appears to scale with the duration of the
GRB ($t_{\rm delay}\sim$ several seconds in the long GRBs 080916C and
090902B, while $t_{\rm delay} \sim 0.1-0.2\;$s in the short GRBs
090510 and 081024B, though with a smaller significance for the
latter).

Only 3 LAT GRBs so far have shown clear ($>\,$5$\,\sigma$) evidence
for a distinct spectral component. However, these GRBs are the 3
brightest in the LAT, while the next brightest GRB in the LAT
(080916C) showed a hint for an excess at high energies. This suggests
that such a distinct high-energy spectral component is probably very
common, but we can clearly detect it with high significance only in
particularly bright LAT GRBs, since a large number of LAT photons is
needed in order to detect it with $>\,$5$\,\sigma$ significance.

\begin{figure}
\centerline{\includegraphics[width=0.7\textwidth]{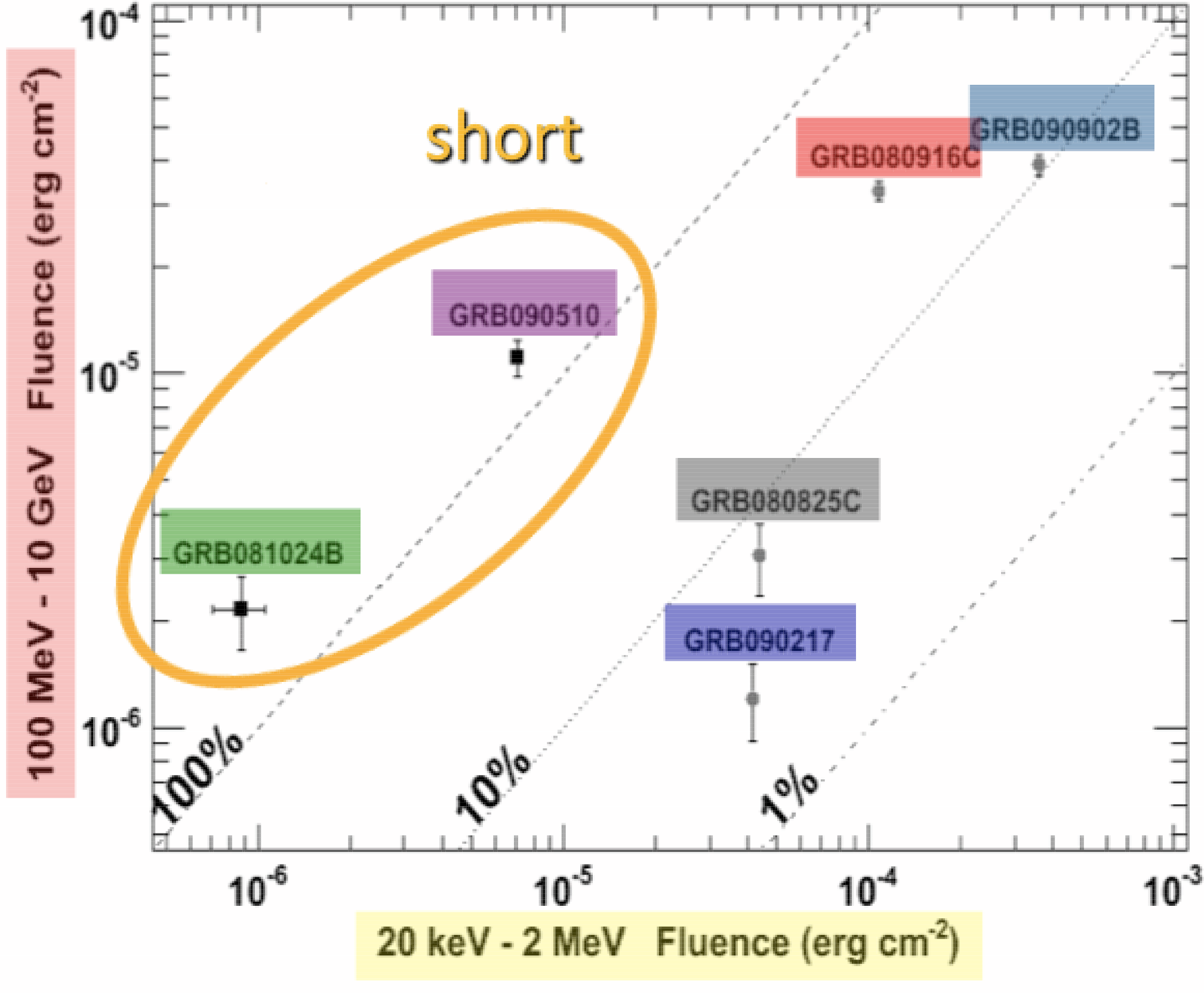}}
\vspace{0.25cm}
\centerline{\includegraphics[width=0.7\textwidth]{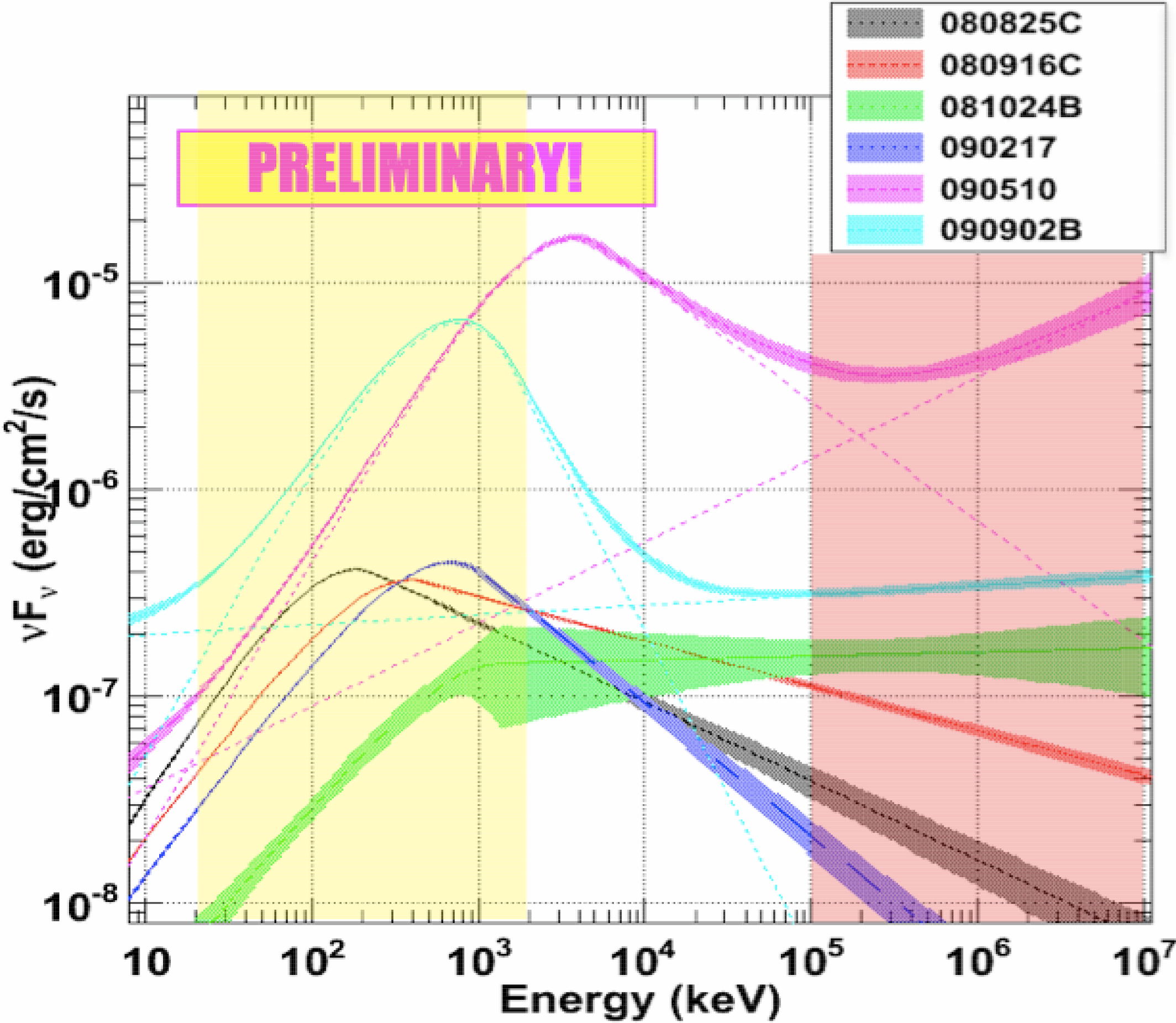}}
\caption{
{\bf Top panel}: the fluence at high (0.1$\,$--$\,$10$\;$GeV) versus
low (20$\,$keV -- $2\,$MeV) energies (from~\cite{081024B}), for 4 long
(080825C, 080916C, 090217, 090902B) and 2 short (081024B, 090510)
duration LAT GRBs. The diagonal lines indicate high to low energy
fluence ratios of 1\%, 10\%, and 100\%. {\bf Bottom panel}: the best
fit time-integrated $\nu F_\nu$ spectra for the same GRBs, two of
which (090510, 090902B) show a distinct spectral component, well
described by a hard power-law, in addition to the usual Band spectral
component. The colored shaded regions indicate the energy ranges used
for calculating the fluences that are displayed in the {\it top
panel}.}
\label{fig:long-vs-short}
\end{figure}

The distinct spectral component is usually well fit by a hard
power-law that dominates at high energies. In GRB~090902B a single
power-law component dominates over the usual Band component both at
high energies (above $\sim 100\;$MeV) and low energies (below $\sim
50\;$KeV; see {\it lower panel} of Fig.~\ref{fig:long-vs-short}).
There is also marginal evidence that the high-energy power-law
component in GRB~090510, which dominates above $\sim 100\;$MeV, might
also appear at the lowest energies (below a few tens of keV).

Both the delayed onset and distinct spectral component relate to and
may help elucidate the uncertain prompt GRB emission mechanism. The
main two competing classes of models are leptonic and hadronic origin.

\noindent
{\bf Leptonic:} the high-energy spectral component might be
inverse-Compton emission, and in particular synchrotron-self Compton
(SSC) if the usual Band component is synchrotron. In this case,
however, it may be hard to produce the observed $t_{\rm delay} > t_v$,
where $t_v$ is the width of individual spikes in the lightcurve
($t_{\rm delay} \lesssim t_v$ might occur due to the build-up of the
seed synchrotron photon field in the emitting region over the
dynamical time). Moreover, the gradual increase in the photon index
$\beta$ of the distinct high-energy power-law spectral component is
not naturally expected in such a model, and the fact that it is
different than the Band low-energy photon index as well as the excess
flux (above the Band component) at low energies are hard to account
for in this type of model.

\noindent
{\bf Hadronic:} $t_{\rm delay}$ might be identified with the
acceleration time, $t_{\rm acc}$, of protons (or heavier ions) to very
high energies (at which they loose much of their energy on a dynamical
time, e.g. via proton synchrotron~\cite{Razzaque09}, in order to have
a reasonable radiative efficiency). If the observed high-energy
emission (and in particular the distinct high-energy spectral
component) also involves pair cascades (e.g. inverse-Compton emission
by secondary $e^\pm$ pairs~\cite{Asano09} produced in cascades
initiated by photo-hadronic interactions) then it might take some
additional time for such cascades to develop. Such an origin for
$t_{\rm delay}$ ($\sim t_{\rm acc}$), however, requires the
high-energy emission to originate from the same physical region over
times $> t_{\rm delay}$, and implies high-energy emission rise and
variability times $t_v \gtrsim t_{\rm acc} \sim t_{\rm delay}$, due to
the stochastic nature of the acceleration process (while $t_v < t_{\rm
delay}$ is usually observed).  The gradual increase in $\beta$ is not
naturally expected in hadronic models, though it might be mimicked by
a time-evolution of a high-energy Band-like spectral
component~\cite{Razzaque09}. For GRB~090510 a hadronic model requires
a total isotropic equivalent energy $> 10^2$ times larger than that
observed in gamma-rays~\cite{Asano09}, which may pose a serious
challenge for the progenitor of this short GRB. The excess flux at low
energies that is observed in GRB~090902B (and the hint for such an
excess in GRB~090510) may be naturally explain in this type of model
by synchrotron emission from secondary pairs~\cite{Asano09,090902B}.

Altogether, hadronic models seem to fare somewhat better, however both
leptonic and hadronic models still face many challenges, and do not
yet naturally account for all of the Fermi observations.

\section{Long-lived high-energy emission}
\label{sec:long-lived}

In most LAT GRBs the high-energy ($>\,$100$\;$MeV) emission lasts
significantly longer than the low-energy ($\lesssim 1\;$MeV) emission.
While the high-energy emission usually shows significant variability
during the prompt (low-energy) emission phase, in some cases showing
temporal correlation with the low-energy emission, the longer lived
emission is typically temporally smooth and consistent with a
power-law flux decay (of $\sim t^{-1.2}-t^{-1.5}$) with a LAT photon
index corresponding to a roughly flat $\nu F_\nu$.

It is most natural to interpret the prompt high-energy emission as the
high-energy counterpart of the prompt soft gamma-ray emission, from the
same emission region, especially when there is temporal correlation
between the low and high-energy light curves, and sometimes even from
the same spectral component (as appears to be the case for
GRB~080916C). However, when there is no such temporal correlation, an
origin from a different emission region is possible. The longer lived
smooth power-law decay phase is more naturally attributed to the
high-energy afterglow, from the forward shock that is driven into the
external medium. An afterglow origin has been suggested in some cases
for the whole LAT emission~\cite{KB-D09,Ghisellini10}, including
during the prompt soft gamma-ray emission stage, however in this
scenario it is generally hard to explain the sharp spikes in the LAT
lightcurve during the prompt phase. It is easier to test the origin of
this long lived high-energy emission when there is good
multi-wavelength coverage of the early afterglow emission (e.g., in
X-ray and/or optical), such as for GRB~090510~\cite{090510-AG}.
Producing particularly high-energy photons is challenging for a
synchrotron origin, both during the prompt emission~\cite{080916C},
and even more so during the afterglow (e.g.~\cite{LW06}), as it
requires a very high bulk Lorentz factor and upstream magnetic field,
in addition to a very efficient shock acceleration (e.g. a $33\;$GeV
photon observed in GRB~090902B after $82\;$s, well after the end of
the prompt emission~\cite{090902B}, requires $\Gamma > 1500$).

\section{High-energy emission of long versus short GRBs}
\label{sec:long-vs-short}

So far, 2 (12) out of the 14 LAT GRBs are of the short (long) duration
class. This implies that $\sim 14\%$ of LAT GRBs are short, with a
large uncertainty due to the small number statistics, which is
consistent with the $\sim 20\%$ short GRBs detected by the GBM. As can
be seen from Table~\ref{tab:LAT14GRBs}, the high-energy emission
properties of short and long GRBs appear to be rather similar. They
can both produce very bright emission in the LAT energy range (090510
vs. 080916C, 090902B and 090926), with a correspondingly high
lower-limit on the bulk Lorentz factor ($\Gamma_{\rm min} \sim 10^3$),
as well a distinct spectral component (090510 vs. 090902B and
090926). Both show a delayed onset and loner lived high-energy
emission, compared to the low-energy emission. However, the delay in
the onset of the high-energy emission appears to roughly scale with
the duration of the GRB, being $\sim 0.1-0.2\;$s for short GRBs and
several seconds for long GRBs. This is especially intriguing when
comparing GRBs 080916C and 090510, which had a comparable isotropic
equivalent luminosity (of several $10^{53}\;{\rm erg\,\,s^{-1}}$),
suggesting another underlying cause for the difference in the time
delay (e.g.~\cite{Toma10}).

Another interesting potential difference, which still needs to be
confirmed (as there are only 2 short LAT GRBs so far, and possible
selection effects), is that short GRBs appear to have a comparable
energy output at high and low photon energies, while long GRBs tend to
radiate a smaller fraction of their energy output at high photon
energies (see {\it upper panel} of Fig.~\ref{fig:long-vs-short}).

 
\noindent
JG gratefully acknowledges a Royal Society Wolfson Research Merit Award.

\end{document}